\documentclass[11pt,cite,a4paper]{article}
\pdfoutput=1
\usepackage{amsmath}
\usepackage{amssymb}
\usepackage{a4wide}
\usepackage{amssymb}
\usepackage{graphicx}
\usepackage{mathrsfs}

\newcommand{\be}{\begin {equation}}
\newcommand{\ee}{\end {equation}}
\newcommand{\ba}{\begin{eqnarray}}
\newcommand{\ea}{\end{eqnarray}}

\def\Im{\mathop{\rm Im}}

\numberwithin{equation}{section}

\begin{document}
 \vspace{2truecm}

\centerline{\Large \bf Fermi Surface evolution under Magnetism Instability}

\vspace{2.3truecm}


\centerline{
    {Yushu Song\footnote{Electric address: yssong2011@gmail.com} and Shu-Qing Guo }
    }
\vspace{.8cm} \centerline{{\it College of Physical Science and Technology, Hebei University, Baoding 071002, China}}
\vspace*{2.0ex} 

\vspace{2.5truecm}

\centerline{\bf ABSTRACT}
\vspace{.5truecm}
In this paper, we study the fermionic excitations near the quantum criticality using gauge/gravity duality. This is motivated by exploring the Fermi surface evolution near the quantum criticality. We construct the gravity dual of ``paramagnetic-nematic" phase transition in a continuum limit and study the Fermi surface evolution across this quantum phase transition. We find that there exists non-Fermi liquid with the Fermi surface in the ``paramagnetic" phase and the Fermi surface disappears in the ``nematic" phase.

 \noindent

\newpage

\section{Introduction}
The traditional condensed matter physics is based on two corner-stones, namely Landau's Fermi liquid theory and Landau's symmetry-breaking theory. Landau Fermi liquids are controlled by a free Fermi gas fixed point with almost no relevant perturbations. While normal state of high-$T_C$ superconducting cuprates and metals close to quantum criticality are examples of non-Fermi liquids whose low energy properties are different from those predicted from Landau's Fermi liquid theory (see~\cite{stewart}). In Landau's symmetry-breaking theory, thermal phase transitions are well developed and quantum phase transitions, which happen at zero temperature from tuning non-thermal control parameters, are traditionally formulated within Landau's paradigm. The proper description of the physics of strongly interacting fermions has long been a major issue for our understanding of condensed matter system near quantum criticality. The gauge/gravity duality has opened new avenues for studying strongly-coupled many body phenomena by relating a classical gravity in a weakly curved (d+1)-dimensional anti-de Sitter spacetime to a strongly-coupled d-dimensional quantum field theory living on its boundary. In particular, black holes have played a universal role in characterizing quantum phases.

In this paper, we study the fermionic excitations near the quantum criticality using gauge/gravity duality. This is motivated by exploring the Fermi surface behavior coupled to the quantum criticality. Following the procedure of \cite{Iqbal:2010eh} \cite{Liu:2009dm}, we construct the gravity dual of ``paramagnetic-nematic" phase transition in a continuum limit and study the Fermi surface evolution across the quantum phase transition. At finite temperature, we study the condensation of Yang-Mills field in the probe limit. At zero temperature limit, we consider the backreaction of Yang-Mills fields and find a fully back-reacted solitonic solution to this system. Then we explore the non-Fermi liquid behavior in these backgrounds. We find that there exists non-Fermi liquid with the Fermi surface in the ``paramagnetic" phase and the Fermi surface disappears in the ``nematic" phase.

An overview of the paper is as follows: In section 2, we set up the holographic model dual to paramagnetic-nematic phase transition. In section 3, we first study the system in the probe limit to determine its phase structure and then we work out the gravity solutions which are dual to ``paramagnetic" and ``nematic" phases in zero temperature limit with backreaction. The former is the closed-form solution, and the latter is numerical. In section 4, we analyze the Fermi surface behavior in different backgrounds. We conclude with a brief discussion of our results and future directions in section 5. In appendix A, we give some details about numerical calculation and show some diagrams.

Note that when we use the word paramagnetic-nematic phase transition there is no sense in which the microscopic degrees of freedom of our system consist of spins that are randomly oriented or aligned on a bipartite lattice in one direction. We only use the term to describe the symmetry breaking pattern. In the discussion of phase transition we are mainly interested in physical scenarios and dynamical mechanisms which arise from the holographic systems rather than detailed phase structure. The discussion of the non-Fermi liquid behavior is similar, and we will not restrict to any specific theory. Since Einstein gravity coupled to matter fields captures universal features of a large class of field theories with a gravity dual \cite{Faulkner:2009wj}, we work with this universal sector. Thus we take the so-called ``bottom-up" approach, i.e. we just consider a certain type of field spectrum without referring to the specific theory.  We expect that if the gravity dual could be constructed top-down for such systems they would likely contain ingredients similar to those in our description.

\section{Set-up of the holographic model}
In a recent paper \cite{Iqbal:2010eh}, the authors constructed a holographic model to study the quantum phase transition and they focused on realizing antiferromagnetic phase where the $SU(2)$ symmetry is broken down to $U(1)$ by the presence of a finite electric charge density. In many models of condensed matter physics, the spin rotations decouple from spatial rotations at long distances, and can be considered as a global symmetry when discussing the low energy physics. In the holographic context, such a global symmetry is described by an $SU(2)$ gauge field in bulk. In this paper, we study a holographic model to realize nematic phase in which the global $SU(2)$ symmetry is broken as well as the rotation symmetry by the presence of current density. In the  bulk, we consider $SU(2) \times U(1)$ gauge theory coupled to AdS gravity as the start point. The condensation of $SU(2)$ gauge field breaks $SU(2)$ gauge symmetry as well as the rotation symmetry. The bulk gauge fields are dual to the conserved currents in the boundary theory. The $U(1)$ gauge field is dual to the boundary chemical potential which does not play an essential role in our discussion.

The bulk theory is described by AdS gravity coupled to Yang-Mills field and Maxwell field with the Lagrangian
 \ba \label{YMMElag}
\mathcal{ L}= \mathcal{R} + \frac{6}{R^2}- {1 \over 4 } F_{MN}^{ a} F^{MN a} - {1 \over 4 } G_{MN} G^{MN},
 \ea
where $\mathcal{R}$ is Ricci scalar, $R$ is the curvature radius of AdS$_4$, $F_{MN}^{ a}$ is Yang-Mills field strength, and $G_{MN}$ is Maxwell field strength satisfying
\ba
&& F_{MN}^a = \partial_M A_N^a - \partial_N A_M^a + iq[A_M, A_N]^a, \cr
&& G_{MN} = \partial_M D_N - \partial_N D_M.
\ea
We will restrict attention to the following ansatz for gauge fields,
\ba \label{ansatzAD}
&& A_M=A(r) \tau^3dt+B(r)\tau^1dx, \\
&& D_\mu=D(r)dt.
\ea
where $q$ is the Yang-Mills coupling and $\tau$ is defined as $\tau^a=\frac{\sigma^a}{2}$ ($\sigma^a$ are the Pauli matrices), satisfying $[\tau^a, \tau^b]=i \epsilon^{abc} \tau^c$. It is straightforward to find the nonzero components of Yang-Mills field strength,
\ba
F^1_{rx}=-F^1_{xr}=B'(r),~~~F^3_{rt}=-F^3_{tr}=A'(r),~~~
F^2_{xt}=-F^2_{tx}=qAB.
\ea
and the nonzero components of Maxwell field strength,
\ba
G_{rt}=-G_{tr}=D'(r).
\ea
According to the anisotropy of Yang-Mills field ansatz in spatial direction, we choose the following ansatz for our metric,
\be \label{ansatzg}
ds^2=-g(r)e^{-\chi(r)}dt^2+\frac{dr^2}{g(r)}+r^2\left(c(r)^2dx^2+dy^2\right).
\ee
With the above assumption, the equations of motion of Yang-Mills field can be reduced to
\ba \label{yme}
A'' +\left(\frac{2}{r}+\frac{\chi'}{2}+\frac{c'}{c} \right) A'-\frac{q^2B^2}{r^2 c^2 g } A=0 \nonumber\\
B'' +\left(\frac{g'}{g}-\frac{\chi'}{2}-\frac{c'}{c} \right) B'+\frac{e^{\chi}q^2A^2}{g^2} B=0
\ea
and the equation of motion of Maxwell field is given by
\be \label{me}
D'' +\left(\frac{2}{r}+\frac{\chi'}{2}+\frac{c'}{c} \right) D'=0
\ee
We use the definition of energy momentum tensor as
\ba
T_{\mu\nu}=-\frac{1}{\sqrt{-g}}\frac{\delta S_{\rm matter}}{\delta g^{\mu\nu}}
\ea
and the nonzero components of energy-momentum tensor are given by
\ba
T_{tt}&=&\frac{1}{4}g A'^2 + \frac{g^2}{4r^2c^2}e^{-\chi}B'^2+\frac{1}{4r^2c^2}\left(q A B\right)^2+\frac{1}{4}g D'^2 \\
T_{rr}&=&-\frac{e^{\chi}}{4g} A'^2 + \frac{1}{4r^2c^2}B'^2+\frac{e^{\chi}}{4g^2 r^2 c^2 }\left(q A B\right)^2-\frac{e^{\chi}}{4g} D'^2 \\
T_{xx}&=&\frac{1}{4}e^{\chi} r^2 c^2 A'^2 + \frac{g}{4}B'^2-\frac{e^{\chi}}{4g}\left(q A B\right)^2+\frac{1}{4}e^{\chi} r^2 c^2 D'^2 \\
T_{yy}&=&\frac{1}{4}e^{\chi} r^2 A'^2 - \frac{g}{4 c^2}B'^2+\frac{e^{\chi}}{4g c^2}\left(q A B\right)^2+\frac{1}{4}e^{\chi} r^2 D'^2
\ea
where $T_{xx}\neq T_{yy}$ which is due to our anisotropic ansatz. Put the above result into the Einstein equations
\be
R_{ab}-\frac{1}{2} g_{ab}\mathcal{R}-3g_{ab}=T_{ab}
\ee
and after a little algebraic manipulation the Einstein equations can be reduced to
\ba
\label{ree}
-\frac{{\chi}'}{r}+\frac{c'}{c}\left(\frac{g'}{g} -{\chi}'\right)&=&\frac{e^{\chi}(q A B)^2}{r^2 g^2 c^2} \nonumber \\
c c''+c c'\left(\frac{g'}{g}+\frac{2}{r} -\frac{{\chi}'}{2} \right)&=&-\frac{B'^2}{2 r^2}+\frac{e^{\chi}}{2 g^2 r^2}(q A B)^2 \nonumber \\
-g'\left(\frac{1}{r}+\frac{c'}{2c}\right)- g \left(\frac{1}{r^2}+\frac{3c'}{c r}+\frac{c''}{c} \right)+3 &=& \frac{e^{\chi}}{4}A'^2+\frac{g B'^2}{4 r^2 c^2}+\frac{e^{\chi} (q A B)^2}{4 g r^2 c^2}+\frac{e^{\chi}}{4 }D'^2 \nonumber \\
\ea
It is to be noted that the above equations (\ref{ree}) have the following scaling invariance
\begin{eqnarray}
\label{rescale}
& &r \rightarrow a_1 r, \quad (t,x,y) \rightarrow (t,x,y)/a_1, \quad g \rightarrow a_1^2g, \quad A \rightarrow a_1 A, \quad B \rightarrow a_1 B, \quad D \rightarrow a_1 D. \nonumber\\
& &e^\chi \rightarrow a_2^2 e^\chi, \quad t \rightarrow a_2 t, \quad A \rightarrow A/a_2, \quad D \rightarrow D/a_2. \nonumber \\
& & x \rightarrow x/a_3,\quad B \rightarrow a_3B. \quad c \rightarrow a_3 c.
\end{eqnarray}
The second scaling symmetry can be used to set $\chi=0$ at infinity and the third one can be used to set $c=1$ at infinity. Thus the metric solution has the asymptotic $\rm {AdS}_4$ form. The fields have the following asymptotical behavior
\ba
A=\mu_A-\frac{\rho_A}{r},\nonumber \\
B=b_0-\frac{b_1}{r}.
\ea
where $\mu_A$ is the chemical potential and $\rho_A$ is the charge density in the boundary theory. In what follows we will only consider the solutions for the field B which vanishes near the boundary, i.e. $b_0 = 0$.

\section{Solution to the holographic model}
\subsection{Phase structure in the probe limit}
In this subsection, we will determine the phase structure of action (\ref{YMMElag}) in the probe limit approximately. The phase diagram of similar action with different boundary condition where one direction is compactified to a circle (and fermions are antiperiodic around this circle) has been discussed in a recent paper \cite{Akhavan:2010bf}. Compared their compactification model, the phase structure of our model is really simple which will be shown below. In order to study the phase structure analytically, we work in the probe limit where the Yang-Mills coupling is large and the backreaction of Yang-Mills on metric is suppressed. It captures most of interesting physics of the phase transition since the nonlinear interactions between Yang-Mills fields are retained. We start with Lagrangian (\ref{YMMElag}), and then rescale Yang-Mills field as $A_M \rightarrow \frac{1}{q} A_M$. The Lagrangian (\ref{YMMElag}) becomes \footnote{We restore the AdS radius R for dimension analysis only in this subsection.}
 \ba \label{YMMElag2}
\mathcal{ L}= \mathcal{R} + \frac{6}{R^2}- {1 \over 4q^2 } F_{MN}^{ a} F^{MN a} - {1 \over 4 } G_{MN} G^{MN}
 \ea
 where the definition of Yang-Mills field strength becomes
 \ba
F_{MN}^a = \partial_M A_N^a - \partial_N A_M^a + i[A_M, A_N]^a
\ea
In the large $q$ limit, we can treat Yang-Mills as the probe and the background is determined by Einstein-Maxwell theory
 \ba \label{YMMElag3}
\mathcal{ L}_{\rm{em}}= \mathcal{R} + \frac{6}{R^2} - {1 \over 4 } G_{MN} G^{MN}.
 \ea
In the following, we will ignore the backreaction of Yang-Mills fields. It would be nice to include the backreaction, at least perturbatively to leading order. The Yang-Mills action is
\ba
\label{ymonly}
S_{\rm{ym}}=-\frac{1}{4q^2}\int d^4x\sqrt{-g} F_{MN}^{ a} F^{MN a}.
\ea
The analytic solution to Einstein-Maxwell theory is  Reissner-Nordstrom-AdS black hole
\ba
\label{ansatzgp}
ds^2=-\frac{g(r)}{R^2}dt^2+\frac{R^2dr^2}{g(r)}+\frac{r^2}{R^2}(dx^2+dy^2),
\ea
with
\ba
D(r)=\mu(1-\frac{r_H}{r}),
\ea
\ba
g(r)=r^2 f(r),~~\eta=\frac{\mu^2}{4}
\ea
\ba
f(r)=1-\frac{r^3_H}{r^3}(1+\eta)+\frac{r^4_H}{r^4}\eta
\ea
The Yang-Mills fields in the  Reissner-Nordstrom-AdS black hole background reduce to
\ba \label{ymeAp}
A'' +\frac{2}{r}A'-\frac{R^4B^2}{r^4 g } A=0 \\
\label{ymeBp}
B'' +\left(\frac{f'}{f}+\frac{2}{r} \right) B'+\frac{R^4A^2}{g^2r^4} B=0
\ea
The above Yang-Mills equations have the following scale symmetry
\ba
r \rightarrow a r, \quad (t,x,y) \rightarrow (t,x,y)/a_1, \quad g \rightarrow g, \quad A \rightarrow a A, \quad B \rightarrow a B.
\ea
Following the procedure of \cite{Gregory:2009fj} we will study the condensation of $B$ field analytically. By changing to the variable $z=\frac{r_H}{r}$,  equations (\ref{ymeAp}) and (\ref{ymeBp}) become
\ba \label{ymeApz}
&A'' -\frac{R^4B^2}{r^2_H f } A=0 &\\
\label{ymeBpz}
&B'' + z^2 \frac{4\eta z -3(1+\eta)}{1+z^4\eta-z^3(1+\eta)}  B'+\frac{R^4A^2}{f^2r_H^4} B=0&
\ea
with
\ba
f(z)=1+z^4\eta-z^3(1+\eta)
\ea
where $z=0$ corresponds to the boundary and $z=1$ corresponds to the horizon.  Next we consider the boundary condition with these variables. Regularity at horizon requires
\ba \label{bcp1}
A(1)=0,~~~B'(1)=0
\ea
and the asymptotic solution near the boundary reads
\ba
A(z)=\alpha_0+\alpha_1z,~~~B(z)=\beta_0+\beta_1z
\ea
We fix the charge density $-r_H\alpha_0=\rho_A$ and take $\beta_0$ to be zero. Around horizon $z=1$, we expand $A(z)$ and $B(z)$ as
\ba
A(z)=A(1)+A'(1)(z-1)+\frac{1}{2}A''(1)(z-1)^2+...\\
B(z)=B(1)+B'(1)(z-1)+\frac{1}{2}B''(1)(z-1)^2+...
\ea
From the boundary condition (\ref{bcp1}), $A(1)=0$ and $B'(1)=0$, and we set $A'(1)<0$ and $B(1)>0$ for positivity of $A(z)$ and $B(z)$. From the equation of motion $A''(1)$ can be expressed as
\ba
A''|_{z=1}=\frac{R^4B^2}{fr_H^2}A|_{z=1}=\frac{R^4B(1)^2 A'(1)}{r_H^2(\eta-3)}
\ea
giving
\ba
A(z)=A'(1)(z-1)+\frac{R^4B(1)^2 A'(1)}{2r_H^2(\eta-3)}(z-1)^2
\ea
The 2nd derivative of $B(z)$ can be calculated similarly as
\ba
B''|_{z=1}&=&-z^2\frac{4\eta z -3(1+\eta)}{1+z^4\eta-z^3(1+\eta)}  B'|_{z=1}-\frac{R^4A^2}{f^2r_H^4} B|_{z=1}\\
&=&-B''|_{z=1}-\frac{R^4B(1)A'(1)^2}{r_H^2(\eta-3)^2}
\ea
Thus we have
\ba
B(z)=B(1)-\frac{R^4B(1)A'(1)^2}{4r_H^2(\eta-3)^2}(z-1)^2
\ea
Around boundary $z=0$, we expand A and B as
\ba
&A(z)=\alpha_0+\alpha_1z+\alpha_2z^2...&\\
&B(z)=\beta_1z+\beta_2z^2...&
\ea
where we have used $\beta_0=0$. Using equation of motion, the 2nd derivatives of A  and B are given by
\ba
&\alpha_2=A''|_{z=0}=0&\\
&\beta_2=B''|_{z=0}=0&
\ea
Then solutions around boundary become
\ba
&A(z)=\alpha_0+\alpha_1z&\\
&B(z)=\beta_1z&
\ea
where $r_H\alpha_1$ is fixed. In order to connect the solutions smoothly, we require the following conditions:
\ba\label{c1}
&\frac{1}{2}\beta_1=b-\frac{R^4ba^2}{16r_H^2(\eta-3)^2}&\\ \label{c2}
&\beta_1=\frac{R^4ba^2}{4r_H^2(\eta-3)^2}&\\ \label{c3}
&\alpha_0+\frac{1}{2}\alpha_1=-{1\over2}a +\frac{R^4b^2 a}{8r_H^2(\eta-3)}&\\ \label{c4}
&\alpha_1=a -\frac{R^4b^2 a}{2r_H^2(\eta-3)}&
\ea
where $A'(1)\equiv a$ and $B(1)\equiv b$ with ($a<0,~~b>0$). From (\ref{c1}) and (\ref{c2}) we can deduce
\ba
\beta_1=\frac{4}{3}b
\ea
Using (\ref{c3}) and (\ref{c4}) we have
\ba
b^2=\frac{2r_H^2(\eta-3)(a-\alpha_1)}{R^4a}
\ea
From the AdS/CFT dictionary, the expectation value of dimension 2 operator is given by
\ba
<\mathcal{O}_2>=\beta_1r_H/R^2
\ea
Using the relation $\rho_A=-r_H\alpha_1$ and Hawking temperature $T=\frac{r_H}{4\pi R^2}(3-\eta)$, $<\mathcal{O}_2>$ can be expressed as
\ba
<\mathcal{O}_2>=\frac{8r_H \sqrt{2\pi r_H}}{3R^3} \frac{T_{\rm{c}}}{\sqrt{T}}\sqrt{1-\frac{T^2}{T_{\rm{c}}^2}}
\ea
where $T_{\rm{c}}$ is defined as
\ba
T_{\rm{c}}=\frac{\sqrt{\sqrt{3}\rho_A(3-\eta)}}{8\pi R}
\ea
We see that $<\mathcal{O}_2>$ is zero at $T=T_{\rm{c}}$ which is a critical point, and the condensation occurs at $T<T_{\rm{c}}$. The critical temperature is the function of $\eta$, so we can lower the critical temperature by adjusting the $U(1)$ charge density. $T_{\rm{c}} \rightarrow 0$ as $\eta\rightarrow 3$, leads to a quantum critical point. Note that compared with \cite{Akhavan:2010bf}, the phase structure of our model is simple. We have only two kinds of solutions, the Reissner-Nordstrom-AdS black hole and a haired black solution. There is a phase transition between them and the critical temperature can be adjusted down to zero.

A few words about the the above analytical calculation are appropriate at this point. Firstly our analytical treatment can be thought of as the leading order approximation of Mathematica NDsolve of equations (\ref{ymeAp}) and (\ref{ymeBp}). Secondly, in our setup, we focus on the nonvanishing $B$ field in bulk, which implies the breaking of $SU(2)$ symmetry and rotation symmetry. Our calculation does not depend on specific value of $\alpha_0$. Vanishing $\alpha_0$ corresponds to the spontaneous $SU(2)$-symmetry breaking. Nonvanishing $\alpha_0$ corresponds to the nonzero chemical potential for $SU(2)$ symmetry and the chemical potential will break $SU(2)$ symmetry directly.

Up to boundary counter terms, the free energy of the Yang-Mills theory is determined by the value of action (\ref{ymonly}) evaluated on-shell, which is the difference of free energy between the condensed phase and the uncondensed phase.
\ba
\Delta \Omega=\Omega_{\rm{condensed}}-\Omega_{\rm{uncondensed}}=-T \widetilde{S}
\ea
where $\widetilde{S}$ is given by
\ba
\widetilde{S}= S_{\rm{ymos}}+S_{\rm{bdy}}+S_{\rm{ct}}.
\ea
Using the equation of motion and the explicit form of metric, the on-shell action reduces to
\ba
S_{\rm{ymos}}=\int d^3x \left[-\frac{1}{2} A'A|_{z=0} +\frac{1}{2}\int^1_0 dz \frac{A^2B^2}{f}\right]
\ea
The most efficient way to regulating $S_{\rm{ymos}}$ is working in the grand canonical ensemble where the chemical potential $\mu$ rather than the charge density is fixed. When working in the grand canonical ensemble, we do not need additional boundary terms and counter terms. Following the procedure of \cite{Herzog:2008he}, we can calculate the $\Delta \Omega$ using the above analytic approximation.
\ba
\Delta \Omega/V=-\frac{1}{2}\mu^2-\frac{1}{2}\int^1_0 dz \frac{A^2B^2}{f}
\ea
The analytic expression of $\Delta \Omega$ shows that the free energy of hair black hole solution is always lower than that of Reissner-Nordstrom-AdS black hole solution. The haired black hole is thermodynamically  preferred.

\subsection{Zero temperature limit}
In this subsection, we will study the phase transition at zero temperature limit with backreaction included. First we study the linearized perturbations of the extremal dyonic-like AdS black hole in order to analyze the instabilities of the backgrounds following the strategy of \cite{Horowitz:2009ij} \cite{Basu:2009vv}. The general dyonic-like AdS black hole solution satisfying the constraint that $g_{\mu\nu}A^\mu A^\nu$ is finite at horizon is given by
\ba \label{ymhair}
& c=1,  \qquad \chi=B=0 &\nonumber \\
& g=r^2-\frac{1}{r}(1+\frac{\rho_1^2+\rho_2^2}{4})+\frac{\rho_1^2+\rho_2^2}{4r^2} &\nonumber \\
& D=\rho_1 \left(1-\frac{1}{r} \right) &\nonumber \\
& A=\rho_2 \left(1-\frac{1}{r} \right) &\nonumber \\
\ea
The horizon radius can be set to one using the first scaling symmetry of (\ref{rescale}). The solution (\ref{ymhair}) is the most general background that preserves rotation symmetry. The temperature of dual field theory is given by
\ba
T& = &{\left[g' (e^{-\chi}g)' \right]^{1/2}\over 4\pi}\nonumber \\
  & = &{12-\rho_1^2-\rho_2^2 \over 16\pi}
\ea
In the zero temperature limit, we have $\rho_1^2+\rho_2^2=12$. On the other hand, in order to study the gravity dual of ``paramagnetic" phase, we must set $\rho_2=0$ otherwise $SU(2)$ symmetry will be broken directly by the charge density. Thus in the background due to ordered phase, the Yang-Mills fields are inactive. The near horizon limit of the extremal solution reduces to AdS$_2 \times \mathbb{R}^2$ with a metric,
\be
ds^2=-6(r-1)^2dt^2+\frac{dr^2}{6(r-1)^2}+dx^2+dy^2
\ee
In the near horizon region, the equation of motion of $B$ field reduces to
\be
\partial_{\tilde{r}}^2 B+\frac{2}{\tilde{r}}\partial_{\tilde{r}} B=0
\ee
where the effective mass of $B$ field is $m_{\rm eff}^2=0$ and $\tilde{r}=r-1$. In our case the Breitenlohner-Freedman bound for AdS$_2$ is satisfied:
\ba \label{unstcon}
m_{\rm eff}^2=0 > m_{\rm BF}^2 = -{1\over4}
\ea
so there is no classical instability in this background. But this background bears the quantum instability due to the Schwinger pair production of fermions \cite{Hartnoll:2011fn} \cite{Pioline:2005pf} if the fermion mass is sufficient low. This pair production will lead to neutralisation of the black hole. In the dual field theory, this corresponds to the phase transition
from ordered phase to the disordered one.

When studied holographic superconductor, people have realized that the zero temperature black holes in Einstein-Maxwell-Higgs theory \cite{Gubser:2009cg} \cite{Horowitz:2009ij} and Einstein-Yang-Mills theory \cite{Basu:2009vv}  have the zero horizon size. Motivated by these facts, we assume the same is true for our model. Similar to that of \cite{Horowitz:2009ij} \cite{Basu:2009vv}, we choose the following ansatz near $r\rightarrow 0^+$
\begin{eqnarray}
&A\sim A_0(r),~~D\sim D_0(r)+D_1(r) , ~~B\sim B_0+B_1(r) ,~~\chi\sim \chi_0+\chi_1(r), &\nonumber \\
&g\sim r^2+g_1(r),~~c\sim c_0+c_1(r).&
\nonumber \\
\label{ansatz}
\end{eqnarray}
Terms with subscript 1 are subleading terms which go to zero faster than leading part. Substituting the above ansatz (\ref{ansatz}) into equations of motion of Yang-Mills fields (\ref{yme}), we have
\ba
r^2 \left (r^2 A'_0(r)\right )'=\frac{q^2 B_0^2} {c_0^2} A_0(r)
\Rightarrow A_0(r) \sim e^{(-\frac{\alpha}{r})}, \quad \alpha=q B_0/{c_0},
\ea
where we have used the constraint that $g_{\mu\nu}A^\mu A^\nu$ is finite at horizon. Following the similar procedures, to first order the equations (\ref{yme}) (\ref{me}) and (\ref{ree}) have the following solutions
\begin{eqnarray} \label{bg2}
& A\sim A_0 e^{-\alpha/r}, ~~~B\sim B_0\left(1-\frac{e^{\chi_0}q^2A_0^2}{4\alpha^2}e^{-2\alpha/r}\right),
~~~c\sim c_0\left(1+\frac{e^{\chi_0}A_0^2}{8r^2}e^{-2\alpha/r}\right), & \nonumber\\
& \chi\sim \chi_0-\frac{e^{\chi_0}A_0^2\alpha}{2r^3}e^{-2\alpha/r},~~~g\sim r^2-\frac{e^{\chi_0}A_0^2\alpha}{2r}e^{-2\alpha/r}, ~~~D\sim 0. & \nonumber\\
\label{scaling}
\end{eqnarray}
where we can set $A_0=1$, $\chi_0=0$ and $c_0=1$ by the scaling symmetries (\ref{rescale}). After fixing  $A_0$, $\chi_0$ and $c_0$, there are two parameters in our near horizon solution (\ref{bg2}), namely $\alpha$ and $q$. Using the equations of motion, we can integrate out the full solution numerically from the near horizon solution. Requiring the $B$ component vanish near the boundary, we will get a nearly linear relation between $\alpha$ and $q$ which is similar to that of \cite{Basu:2009vv} and we will show the diagram here. The gravity solution (\ref{bg2}) not only breaks the $U(1)_3$ gauge symmetry generated by $\tau^3$ but also picks out $x$ direction as special. So the background (\ref{bg2}) captures the physics of the nematic phase which breaks the gauge symmetry as well as the rotation symmetry.

It is worth to explain the gauge/gravity duality in our setup a little further here. From the discussion of section 3.1, there is a Hawking-Page phase transition between different vacua in the gravity side. According to the AdS/CFT dictionary \cite{Witten:1998zw}, the Hawking-Page phase transition is dual to the confinement-deconfinement phase transition in field theory, and the black hole is dual to the deconfined phase. It is natural to associate the flux emanating from the horizon with the fractionalized charge carrier. So the paramagnetic phase is partially fractionalized \cite{Hartnoll:2011pp}. In the process of phase transition, the black loses its charge by Schwinger pair production of charged fermionic particles. From the field theory point of view, all ``quarks" are confined into ``mesons". The paramagnetic phase jumps into the nematic phase which is fully mesonic.

Before concentrating on the fermionic probe, we can make some clarifications about gravity solutions here. If we want to find rotation invariance solution of our equations of motion (\ref{yme}) (\ref{me}) and (\ref{ree}), we must set $c=1$ and $B=0$ for our ansatz (\ref{ansatzAD}) and (\ref{ansatzg}). With these conditions, the equations of motion can be carried out with the closed-form solution (\ref{ymhair}). There are two free parameters $\rho_1$ and $\rho_2$ in the solution. In this paper, we work in the zero temperature limit where $\rho_1^2+\rho_2^2=12$. First, when $\rho_2=0$ and $\rho_1=2\sqrt3$, the solution preserve $SU(2)$ gauge symmetry and rotation symmetry around x and y directions. This background captures the physics of the paramagnetic phase and bears the quantum instability due to Schwinger pair production of fermions though there is no classical instabilities. Second, when $\rho_2\neq0$ and $\rho_1^2+\rho_2^2=12$, this solution breaks the $SU(2)$ symmetry down to $U(1)$ but preserves the rotation symmetry. This background is dual to the ferromagnetic phase with external magnetic field nonvanshing since the external magnetic field is dual to the chemical potential of $SU(2)$ gauge field. In this background, there are both classical instability and quantum instability. The classical instability comes from the violation of BF bound by effective mass of Yang-Mills field component $B$. This is consistent with the thermodynamics analysis \cite{Hartnoll:2008kx} where the additional external field raise the free energy. Comparing with the above two solutions, our numerical solution is preferred at quantum level. In the following, we will focus on Fermi surface evolution across the paramagnetic-nematic phase transition. We leave the Fermi surface evolution across the paramagnetic-ferromagnetic phase transition for future work while the discussion about Fermi surface evolution across the ferromagnetic-nematic phase transition is similar.

\section{Fermi surface behavior in different backgrounds}

Now we focus on the effect of the Yang-Mills field condensation on the non-Fermi liquid behavior. The basic idea, introduced in \cite{Liu:2009dm}, is to introduce fermions in the above geometries. The Green's functions of fermions probe the existence and the properties of the fermi surface in the boundary theory. In our models, we suppose the fermions are charged under the $U(1)$ symmetry. The Yang-Mills fields affect the Fermi surface behavior only through background metric.

To compute the spectral functions for fermionic operator in dual field theory, we need only quadratic action of $\psi$ in geometry (\ref{ymhair}) and (\ref{ansatz})
\be \label{dirac}
S_{\rm {spinor}}=\int d^{d+1}x \sqrt{-g} i\left(\bar{\psi} \Gamma^M\mathcal{D}_M \psi-m\bar{\psi}\psi  \right)
\ee
where
\ba
\bar{\psi}=\psi^\dag \Gamma^{\underline{t}},~~~ \mathcal{D}_M=\partial_M+\frac{1}{4} \omega_{abM}\Gamma^{ab}-iq_f D_M,
\ea
and $\omega_{abM}$ is the spin connection. We will use $M, N...$ and $a,b...$ to denote bulk spacetime and tangent space indices respectively, and $\mu, \nu...$ to denote indices along the boundary directions, i.e. $M=(r,\mu)$. Although our discussion does not depend on dimension $d$, we will specify $d=3$ which is dual to 2+1 dimensional field theory in the boundary. To analyze the Dirac equations following from (\ref{dirac}), it is convenient to choose the following basis
\ba
 \label{gct}
\Gamma^r = \left( \begin{array}{cc}
1 & 0  \\
0 & -1
\end{array} \right), \;\; \Gamma^\mu = \left( \begin{array}{cc}
0 & \gamma^\mu  \\
\gamma^\mu & 0
\end{array} \right), \;\; \psi = \left( \begin{array}{c} \psi_+ \\ \psi_- \end{array}\right)
\ea
where $\psi_\pm$ are the two-component spinors and $\gamma^\mu$ are (2+1) dimensional gamma matrices. We can seperate the radial and boundary coordinate dependencies in $\psi$ as follows
\ba \label{psiexp}
\psi_\pm =  (- g g^{rr})^{-{1 \over 4}} e^{-i \omega t + i k_i x^i} \phi_\pm \ ,~~~\phi_\pm = \left( \begin{array}{c}
y_\pm  \\
z_\pm
\end{array} \right)
 \ea

First, we consider Dirac equation in the dyonic-like AdS black hole background (\ref{ymhair}) in the zero temperature limit since we are interested in quantum phase transition. The Dirac equation from the quadratic action is
\ba
\Gamma^M \mathcal{D}_M \psi-m\psi=0.
\ea
The background metric can be written as
\ba
&ds^2=-g(r)dt^2+\frac{dr^2}{g(r)}+r^2(dx^2+dy^2),& \nonumber\\
&g(r)=r^2-\frac{1}{r} \left(1+\frac{\rho_1^2+\rho_2^2}{4} \right)+\frac{\rho_1^2+\rho_2^2}{4r^2}.&
\ea
The Maxwell field $D$ is given by
\ba
D=D_t dt,~~~D_t=\rho_1\left(1-\frac{1}{r}\right).
\ea
The Yang-Mills field is
\ba
 A_M=A(r) \tau^3dt,~~~ A=\rho_2 \left(1-\frac{1}{r} \right).
\ea
In the zero temperature limit, $\rho_1$ and $\rho_2$ satisfy the following condition:
\ba
\rho_1^2+\rho_2^2=12
\ea
Using the coordinate seperation (\ref{psiexp}), the Dirac equation becomes
\be \label{eropi}
\sqrt{ g_{ii} \over  g_{rr}} \left( \partial_r  \mp m \sqrt{g_{rr}}\right) \phi_\pm
= \mp i K_\mu \gamma^\mu \phi_\mp,
\ee
where
\be \label{uen}
K_\mu (r) = \left(-u(r),  k_i \right), \qquad  u= \sqrt{g_{ii} \over - g_{tt}} (\omega + q_f \rho_1 (1-{1 \over r})) \ .
\ee
We choose the following basis $\gamma^0=i\sigma_2,~~\gamma^1=\sigma_1,~~\gamma^2=\sigma_3$ for the $\gamma^\mu$ and set $k_2=0$ using the rotational symmetry of the dyonic-like AdS black hole background. The Dirac equations can be reduced to the following two sets of decoupled equations
\ba
\label{decoupled}
\sqrt{\frac{g_{ii}}{g_{rr}}} (\partial_r \mp m\sqrt{g_{rr}})y_\pm =
\mp i(k_1 - u)z_\mp, \nonumber\\
\sqrt{\frac{g_{ii}}{g_{rr}}} (\partial_r \pm m\sqrt{g_{rr}})z_\mp =
\pm i(k_1 + u)y_\pm,
\ea
We introduce the ratios $\xi_+ = iy_-/z_+$, $\xi_- = -iz_-/y_+$, in terms of which the above equation (\ref{decoupled}) can be written as
\be
\label{xiEquations}
\sqrt{\frac{g_{ii}}{g_{rr}}} \partial_r \xi_\pm = -2m\sqrt{g_{ii}}\xi_\pm
\mp (k_1 \mp u) \pm (k_1 \pm u)\xi_\pm^2.
\ee
The retarded Green's function $G_R$ is given in terms of the quantities $\xi_\pm$ by:
\ba
\label{retardedgf}
G_R = \lim_{\epsilon \rightarrow 0} \epsilon^{-2m}
\left.\begin{pmatrix} \xi_+ & 0 \\ 0 & \xi_- \end{pmatrix}\right|_{r = \frac1\epsilon}
\equiv \begin{pmatrix} G_{11} & 0 \\ 0 & G_{22} \end{pmatrix}
\ea
From now on we will drop the subscript 1 on momentum $k_1$. The spinors $\xi_\pm$ satisfy the in-falling boundary condition at the horizon which implies
\be \label{tye}
\xi_\pm|_{r=1} =i \ .
\ee
At $\omega=0$, the in-falling boundary condition (\ref{tye}) should be replaced by
\be \label{roep}
 \xi_{\pm}|_{r=1, \, \omega =0}= {  m - \sqrt{k^2 + m^2 - {\mu_q^2 \over 6} - i \epsilon} \over {\mu_q \over \sqrt{6}} \pm k} \ .
 \ee
With the boundary conditions, we can integrate equation (\ref{xiEquations}) numerically to $r \to \infty$ obtained the boundary correlation function. The imaginary part of the retarded function $G_R$ is the function of $k$ and $\omega$ for fixed $m$, $q_f$ and $\rho_1$. The equation (\ref{xiEquations}) which gives the location and property of the Fermi surface is nearly the same as that of Liu, McGreevy and Vegh \cite{Liu:2009dm}, so we find the similar Fermi surface behavior as that of Liu et.al. We will not repeat the numerical calculation here(we leave the numerical details in Appendix A), but the conclusion is obvious due to Liu et al.\cite{Liu:2009dm}. For some fixed $\omega < 0$, $\Im G_{22}$ has a sharp peak at some momentum value $k_F$ which implies the Fermi surface exist. Also from the scaling behavior of $\Im G_{22}$, it is different from that of Landau Fermi liquid where $z=\alpha=1$ in
\ba
\Im G_{22}(\lambda^z \omega, \lambda k_{\perp})=\lambda^{-\alpha} \Im G_{22}(\omega, k_{\perp})
\ea
In the region $k< \frac{\mu_q}{\sqrt{6}}$, the the limit $\omega \rightarrow 0$, $\Im G_{ii}$ becomes oscillatory with oscillatory peak periodic in $\log|\omega|$ with constant heights which is a new phenomena of non-Fermi liquid. For more details refer to \cite{Liu:2009dm}.
Next we will consider the Dirac equation in the background whose near horizon limit gives (\ref{scaling}). The key difference of this numerical background is that there is no rotation symmetry in the $x-y$ plane. Using the similar technique, the Dirac equation can be reduced to the following
\ba \label{cdirac1}
\left( -\sqrt{g} \partial_r  + m \right) \phi_+
&=& \left[ \frac{e^{\chi/2}}{\sqrt{g} } (-i\omega \gamma^0)+\frac{i k_1 \gamma^1}{r c} +\frac{i k_2 \gamma^2}{r } \right]\phi_-,\\
\label{cdirac2}
\left( \sqrt{g} \partial_r  + m \right) \phi_-
&=& \left[ \frac{e^{\chi/2}}{\sqrt{g} } (-i\omega \gamma^0)+\frac{i k_1 \gamma^1}{r c} +\frac{i k_2 \gamma^2}{r } \right]\phi_+
\ea
Near $r\rightarrow \infty$, $\phi_\pm$ have the following asymptotic behavior,
\ba
\phi_+=Ar^m+Br^{-m-1},~~ \phi_-=Cr^{m-1}+Dr^{-m}
\ea
The coefficients $D$ and $A$ are related by a matrix $\mathcal{S}$,
\be
D=\mathcal{S} A
\ee
and then the retarded Green function $G_R$ is given by \cite{Iqbal:2009fd}
\be \label{GRdef}
G_R=-i\mathcal{S} \gamma^0
\ee
In the near horizon region, the equation (\ref{cdirac1}) (\ref{cdirac2}) becomes coupled equations for $\phi_\pm$
\ba
r(r \partial_r -m)\phi_+=-i \gamma \cdot k \phi_- \\
r(r \partial_r +m)\phi_-=-i \gamma \cdot k \phi_+
\ea
where $\gamma \cdot k=-\omega \gamma^0 +k_1 \gamma^1+k_2 \gamma^2$. These coupled equations can be solved using Bessel functions, and the solution satisfying the in-falling boundary condition at horizon is
\be
\phi_{+} = \begin{cases}
 r^{-{1 \over 2}} K_{m + \frac{1}{2}} \left({\sqrt{|\vec{k}|^2-\omega^2 }\over r} \right) a_+  &  k^2 > 0 \\
 r^{-{1 \over 2}} H^{(1)}_{m+\frac{1}{2}} \left({\sqrt{\omega^2 - |\vec{k}|^2} \over r} \right) a_+  & \omega > |\vec{k}| \\ r^{-{1 \over 2}} H^{(2)}_{m+\frac{1}{2}} \left({\sqrt{\omega^2 - |\vec{k}|^2} \over r} \right) a_+  & \omega < -|\vec{k}| \\
\end{cases}
\ee
where $a_+$ is an arbitary constant spinor.
In the followings, we will focus on the spin averaged spectral function which is defined as
\be
A(\omega,\vec{k})\equiv\frac{1}{\pi} \Im (\rm{Tr} G_R)
\ee
From the numerical solution of equation (\ref{cdirac1}) and (\ref{cdirac2}) satisfying the proper boundary conditions, we can find the behavior of spectral function $A(\omega,\vec{k})$ with parameters $m$ and $q$ freely varied. For definiteness, let us now focus on $m=0$ and $q=1$. In this case the spectral function $A(\omega,\vec{k})$ is a function of $\omega$, $k_1$ and $k_2$. We can plot the three-dimensional diagram of spectral function $A(\omega,\vec{k})$ as the function of two variables of $\omega$, $k_1$ and $k_2$. Using mathematica we find that in the region $-1<\omega<1$, $-3<k_1<3$, and $-3<k_2<3$ the spectral function is regular which implies that the Fermi surface disappears.

The disappearance of Fermi surface can be understood as follows. In the ordered phase, the Reissner-Nordstrom-AdS black hole sources an electric flux towards infinity and the charge-carrying fields in the bulk can be neglect in the large N limit. The fermion spectral function exhibit Fermi surface singularities. The origin of the non-Fermi liquid is from the strong interaction between fermionic excitations and quantum criticality dual to the IR AdS$_2$ geometry. After the phase transition, the black hole loses their charge due to Schwinger pair production and the IR geometry AdS$_2 \times \mathbb{R}^2$ collapses to the zero horizon gravity solution. In this paper, we do gravity calculation classically without including the backreaction of the pair production particles and so the charge density vanishes in our numerical background solution. If the backreaction is included, our numerical background will be modified to something like the electron star \cite{Hartnoll:2010gu}. It is interesting to study the fully back-reacted solution of disordered phase in our model in the future.

\section{Summary and Discussion}
In this paper, we study the fermionic excitations near the quantum criticality in the holographic model. We construct the gravity dual of ``paramagnetic-nematic" phase transition in a continuum limit and study the non-Fermi liquid behavior across the quantum phase transition. We also study the Yang-Mills condensation at finite temperature in the probe limit. We consider the Yang-Mills field in the background of extremal dyonic-like AdS$_4$ black hole and find a fully gravity back-reacted solution to this system at zero temperature limit. The non-Fermi liquid behavior is studied in this background and it is found that the Fermi surface disappears across the quantum phase transition from gravity side. Though the Fermi surface disappears in the ``nematic" phase, it is interesting to explore the non-Fermi liquid behavior further using other techniques. The above discussion on Fermi surface behavior across quantum phase transition can be generalized to other background such as Lifshitz case \cite{Alishahiha:2012nm}. As mentioned earlier, it is also interesting to construct our holographic model from the top down. In that case, we will know more about the field content and interaction form.

In our model, the Yang-Mills fields obtain expectation value spontaneously which is not a standard problem in high energy physics since such an expectation value breaks the Lorentz invariance. Understanding the Fermi surface behavior in some other kinds of quantum phase transition, such as the paramagnetic-ferromagnetic phase transition, is still an interesting open question. One thing we should mentioned is that the gravity background dual to the nematic phase constructed in this paper is only a numerical solution and the analytical solutions for these kinds of backgrounds will be interesting. It would be desirable to study our construction at finite temperature with backreaction included. In that case the quantum phase transition becomes a thermal one and we can compare the holographic model result with the condensed matter result \cite{Yamamoto:2010}.

\section*{Acknowledgments}
We would like to thank Zheng-Xin Liu, Hong Lu, Jian-Huang She, Tower Wang, Yong Xiao and Yang Zhou for helpful discussions. We are grateful to Xiao-Dong Li for kind help on mathematica and the paper would have never seen the light without his help. We also thank the Email correspondence of Mohsen Alishahiha and Hong Liu. This of course does not mean that they are responsible for the opinions and conclusions of this paper. This work was supported by CNSF 11147177.

\section*{Appendix A: Some Numerical Results}
In this appendix, we will explore the location of the Fermi surface in two different backgrounds (\ref{ymhair}) and (\ref{bg2}) using numerical methods. First we solve equation (\ref{xiEquations}) numerically with boundary condition (\ref{tye}) and (\ref{roep}) using mathematica NDsolve, and then we plot $\Im G_{ii}$ using formula (\ref{retardedgf}) which gives the information about spectral function. For definiteness, we focus on $m=0$, $q=1$, and $\rho_1=\sqrt3$ and then our equation (\ref{xiEquations}) is the same as that of Liu, McGreevy and Vegh \cite{Liu:2009dm}. So we find the same diagram as that of Liu et al. We will not repeat it here. The result is that there is a sharp peak in $\Im G_{22}$ showing the singular behavior of spectral function which implies the existence of Fermi surface.

Secondly, about the numerical background which is dual to ``nematic" phase we need to NDsolve equation (\ref{cdirac1}) and (\ref{cdirac2}). Then using the definition of retarded Green's function, we can plot $\Im G_{ii}$
in the numerical background. We plot $\Im G_{ii}$ as function of $k_1$ and $k_2$ while $\omega$ scans some region. We find that in the region $-1<\omega<1$, $-3<k_1<3$, and $-3<k_2<3$ the spectral function is regular which implies that the Fermi surface disappears.

\end{document}